# Tailored Thermal Transport in Phase Change Materials-Based Nanocomposites through Interfacial Structuring

**Viktor Mandrolko[1] and Mykola Isaiev[1]**

**[1]Université de Lorraine, CNRS, LEMTA, F-54000 Nancy, France**

## Abstract

Interfacial thermal transport is a critical bottleneck in nanoscale systems, where heat dissipation and energy efficiency are strongly modulated by molecular ordering at solid–liquid boundaries. Here, using atomistic simulations of hexadecane confined by structured silica substrates, we reveal how interfacial geometry, specifically curvature, governs the density distribution and thermal transport across the interface.

On flat and mildly curved surfaces, the liquid exhibits surface-templated layering, promoting efficient heat transfer, which is enhanced as the contact surface area increases. As curvature increases, this ordering breaks down, giving rise to interference-like density patterns, reduced molecular packing, and localized depletion zones. This structural reorganization leads to a systematic increase of up to 10% in interfacial thermal resistance (ITR), even when the contact area is kept constant.

By decomposing the interface into convex (“hills” of the solid) and concave (“valleys” of the solid) regions, we find that valleys consistently exhibit lower ITR. In contrast, hills act as bottlenecks to heat flow, leading to interfacial thermal resistance values up to 70% higher than those of valleys, depending on the surface configuration. Remarkably, we show that the work of adhesion and entropy-related energy gain upon liquid detachment scale non-trivially with curvature: while adhesion increases with contact area from 30 $mJ \cdot m^{-2}$ for a flat surface to 40 $mJ \cdot m^{-2}$ for a maximum curvaceous surface, the entropic penalty dominates the total energy change, reflecting curvature-induced frustration of molecular alignment.

## 1. Introduction

Due to the increasing consumption of energy resources and the resulting rise in greenhouse gas emissions, enhancing the role of green energy sources has become crucial. One of the major contributors to energy consumption is the heating and cooling of buildings. In this context, phase change materials (PCMs) offer a promising solution. These materials can accumulate thermal energy during periods of excess, store it, and release it when needed, thereby improving energy efficiency and reducing overall consumption.

PCM-based systems are indispensable devices that enhance the efficiency of thermal energy storage from renewable sources, such as solar [1] and geothermal energy [2]. Moreover, PCMs-based storage cells allow the harvesting of industrial waste energy for further re-usage, e.g., maintaining appropriate working conditions [3]. The wide range of applications of PCMs-based thermal storage systems is due to the significant latent heat that can be accumulated and released during the solid/liquid phase transition. However, most known PCMs have a relatively low thermal conductivity (~0.1-0.5 $W \cdot m^{-1} \cdot K^{-1}$), which limits their charging and discharging rates during solidification and melting. This issue can be overcome by using solid nanoparticles [4], [5], or porous medium/solid foam [6], [7] to tailor the thermal transport properties of the composite material. Specifically, previous studies have reported enhancements in the thermal transport properties of the confinement PCMs [8], reductions in the degree of supercooling [9] improvement in photothermal conversion efficiency [10].

The strategies mentioned above for improving the efficiency of PCM-based systems require an accurate understanding of the phenomena that occur at the interface between a PCM and a nanostructured solid. Indeed, changes in properties can be considered in systems with a large interfacial area [10], and the interactions of molecules/atoms near the interface are critical in this respect. For instance, heat and mass transfers are known to be significantly modified at the interfaces between different materials (i.e., solid and liquid ones). Noticeable changes are also reported for properties such as thermal conductivity close to a nanoparticle [12] or at fluid-solid interfaces in a porous network [13], [14]. The latter significantly impacts the effective thermal conductivity of the resulting nanocomposite, given its significant specific surface area [15], [16]. Therefore, the primary strategies for further enhancing the thermal transport properties of composites are the chemical modification of the solid-liquid interface [17], as well as increasing the specific surface area [18], [19]. Nevertheless, the adhesion between the solid and liquid should be considered to prevent leakage [20].

Despite these advances, the influence of molecular distribution and ordering on phase change processes at interfaces remains largely ununderstood. A molecular-level perspective is crucial for accurately describing interfacial phenomena. Atomistic simulation techniques, particularly molecular dynamics (MD), provide a powerful approach to capture both interfacial structure and bulk PCM behavior. Specifically, molecular dynamics (MD) techniques effectively investigate the behavior of various systems based on the interactions between individual atoms. As a result, it allows the study of thermophysical [21], [22], [23], rheological [24], [25], [26], and structural[12], [27] characteristics of PCMs and PCM-based nanocomposites.

Among the numerous types of PCMs, n-alkanes are considered because of their chemical and thermal stability. N-alkanes are widely used in thermal control and energy storage applications. For example, n-alkanes are of interest in the design of hydrocarbon-based superconductors [28] and oleophobic surfaces [29]. The phase transition temperature of n-alkanes is in the range of ambient temperature [30]. Due to its broad use, we focus on the n-alkane with 16 carbon atoms, also called hexadecane ($C_{16}H_{34}$). MD techniques have already been used to investigate hexadecane. It has been shown that the temperature behavior of viscosity, density, and interfacial tension calculated with MD [31] for hexadecane correlates well with experimental data [25], [28], [32], [33]. However, since MD simulations require the specification of a suitable interaction potential and its subsequent parametrization, the application of this methodology to complex composite systems remains limited, with a focus mainly on investigating carbon-based materials and nanocomposites [17], [18].

In practical PCM-based nanocomposites and porous thermal storage media, phase change processes occur predominantly at solid‑liquid interfaces, which are often highly curved due to nanoscale porosity, surface roughness, or deliberate nanostructuring. At these interfaces, surface curvature directly affects molecular packing, orientational ordering, and the efficiency of heat transfer between the PCM and the solid matrix. As a result, the interfacial thermal resistance and local heat flux can vary significantly with surface geometry, thereby influencing the charging and discharging rates of PCM-based systems [34], [35]. Furthermore, the magnitude and trend of interfacial thermal conductance are strongly dependent on interfacial affinity (liophilicity/liophobicity) [36]. Understanding how interfacial curvature governs molecular organization and thermal transport is therefore essential for the rational design of PCM-based composites with enhanced thermal performance. The present study is directly motivated by our experimental investigations of solid‑liquid nanocomposites based on porous silicon and silicon nanowire matrices fabricated by electrochemical and metal-assisted chemical etching. In both systems, forming a solid‑liquid nanocomposite enhances thermal conductivity relative to the corresponding pristine matrices. However, this enhancement is significantly more pronounced for porous silicon with a fractal-like, highly curved morphology, while it is much weaker in silicon nanowire systems that represent smooth interfacial geometries [37], [38]. This opens the possibility of optimizing the thermal transport properties in PCMs-based nanocomposite systems. However, since the well-developed interface is the source of the phonon scattering in nanostructured materials, the compromise between improvement of the thermal properties and interfacial contacting area should be found [39].

In this work, we investigate thermal transport across a nanostructured solid–liquid interface using MD simulations. The solid substrate is amorphous silica patterned with sinusoidal nano-curvatures of varying amplitude and period, and the liquid phase is n-hexadecane. Amorphous silica is a relevant material for PCM encapsulation [29] and is commonly present as an oxide layer in silicon-based porous matrices [14]. We evaluate the interfacial thermal resistance (ITR) across these interfaces, as well as the work of adhesion between hexadecane and the nanostructured silica. This study is distinguished from previous works by examining curved, amorphous solid surfaces with embedded charges (silica) in contact with a long-chain organic PCM. By systematically varying the curvature (through sinusoidal surface features) and quantifying the effects on ITR and adhesion, we aim to uncover molecular-level mechanisms by which surface geometry influences heat transfer. We also decompose the contributions of convex and concave regions of the interface to thermal transport and identify the thermodynamic factors (adhesive vs. entropic) that govern energy dissipation during liquid–solid detachment. Our findings reveal a direct correlation between surface curvature and interfacial transport properties, indicating that nanoscale geometry can be harnessed as a design parameter to tune heat flow in PCM-based nanocomposites.

# 2. Simulation details

## 2.1. The nanostructured solid surface formation

For the simulations of the solid surface, we initially considered the bulk alpha crystalline silica. The lattice was orthogonalized, resulting in lattice parameters of 5.03, 8.71, and 5.52 Å along the *x*-, *y*-, and *z*-directions, respectively. The simulation domain was constructed by replicating 15 × 10 × 30 unit cells in these directions, yielding a system of 81,000 atoms. The ClayFF potential [40] represented the interactions between atoms. In our system only non-bonded interactions are included represented by 12 – 6 Lennard-Jones and Coulomb components. The system was heated to a temperature of 4200 K and relaxed at this temperature for 2 ns in an NPT thermostate. The quenching procedure was performed during 3.9 ns by slow cooling of the system with a cooling rate 1 K/ps [41] up to the temperature equal to 300 K. The final density of the amorphous silicon was to be equal to 2.13 g/cm$^3$, which corresponds well to the literature data [42].

The system was cut into two slabs separated by the sine-like surface set by the equation:

$$z(x) = A \cdot sin\left(2 \cdot \pi \frac{x}{a}\right) + \frac{h_z}{2}, \tag{1}$$

where $A$ is the amplitude, $a$ is the period of the sinusoidal features, and $h_z$ is the vertical offset along the $z$-axis. To ensure compatibility with periodic boundary conditions in the $x$-direction, the period $a$ was set such as $a = \frac{L_x}{\nu}$, where $L_x$ is the simulation box in the $x$-direction and $\nu$ is the number of periods within $L_x$. The amplitude $A$ was set to 2, 4, 6, and 8 Å, with $\nu$ assigned 2, 3, or 4 periods for each amplitude. Additionally, a flat surface was included ($A = 0$, $\nu = 0$), resulting in a total of 13 systems.

Due to the limited box size, directly cutting the system along the sinusoidal surface could result in unequal numbers of atoms in the two slabs, especially near the top and bottom regions of the simulation box. To address this, the system was effectively constructed by uniformly shifting all atomic positions along the $z$-axis by a displacement $h$, taking into account periodic boundary conditions, for fixed values of $A$ and $a$, five distinct values of $h$ were selected to generate two independent, electrically neutral substrates, each comprising an equal number of identical atoms. This methodology enabled the examination of interfacial thermal transport across varying interfaces while maintaining consistent surface morphology and atomic composition.

To quantitatively characterize the geometrical complexity of the corrugated surfaces, the mean curvature of the sinusoidal profile was evaluated. The local curvature $\kappa(x)$ of a curve $z(x)$ can be expressed as:

$$\kappa(x) = \frac{|z''(x)|}{(1 + [z'(x)]^2)^{\frac{3}{2}}}, \tag{2}$$

Where $z'(x)$ and $z''(x)$ denote the first and second derivatives of $z(x)$. The mean curvature $\overline{\kappa}$ was then computed as the spatial average over one period:

$$\overline{\kappa} = \frac{1}{L_x/\nu} \int_{0}^{L_x/\nu} \kappa(x)\, dx \,, \tag{3}$$

This approach allows for systematic comparison of surface curvature across different amplitude–period combinations.

In addition to curvature, the effective interfacial contact area was explicitly quantified. For corrugated surfaces, the effective area was defined as the actual surface area of the sinusoidal interface rather than the projected box area. Specifically, the arc length of the sinusoidal profile $z(x)$ was computed as:

$$L_{eff} = \int_{0}^{L_x} \sqrt{1 + \left(\frac{dz}{dx}\right)^2}\, dx \,, \tag{4}$$

The effective interfacial area was then obtained as $L_{eff} \cdot L_y$, with $L_y$ being the box length in the transverse direction. For the flat surface, this definition naturally reduces to the cross-sectional area ($\Sigma = L_x \cdot L_y$).

## 2.2. Interfacial thermal resistance calculation

To explore how silica surface nanostructuring affects interfacial thermal transport, we employed the non-equilibrium molecular dynamics (NEMD) method. This approach enabled us to measure the heat flux across the interface and calculate the interfacial thermal resistance (ITR). The simulation setup for 4 representative cases is shown in Figure 1. It is consisted of two nanostructured silica slabs separated by a hexadecane layer of 1600 molecules. Periodic boundary conditions were applied in all directions, with 12 Å of empty space along the *z*-axis to prevent interactions between the top and bottom silica slabs via periodic boundaries.

The molecular interactions in hexadecane were modeled using the all-atom optimized L-OPLS force field [43], which provides accurate agreement with experimental structural and thermodynamic properties [44]. This potential includes both bonded (bond stretching, angle bending, and dihedral torsions) and non-bonded interactions, with all parameters taken from the work of Siu et al. [43]. For the interactions between hexadecane and the silica surface, the arithmetic mixing rule was employed to determine the cross-interaction parameters. As a result, the silica–hexadecane interactions were described via a combination of Lennard-Jones and electrostatic potentials. All LJ interactions and the real-space part of the Coulomb interactions were truncated at cutoff equal to 10 Å. Long-range electrostatics were treated with the particle–particle particle–mesh (PPPM) method, with a target relative accuracy of $10^{-4}$ on the electrostatic forces.

After thermalization, the hexadecane region extended approximately 114 Å along the *z*-axis, with the overall simulation box dimensions being 79.2 Å, 91.4 Å, and 310 Å in the *x*, *y*, and *z* directions, respectively.

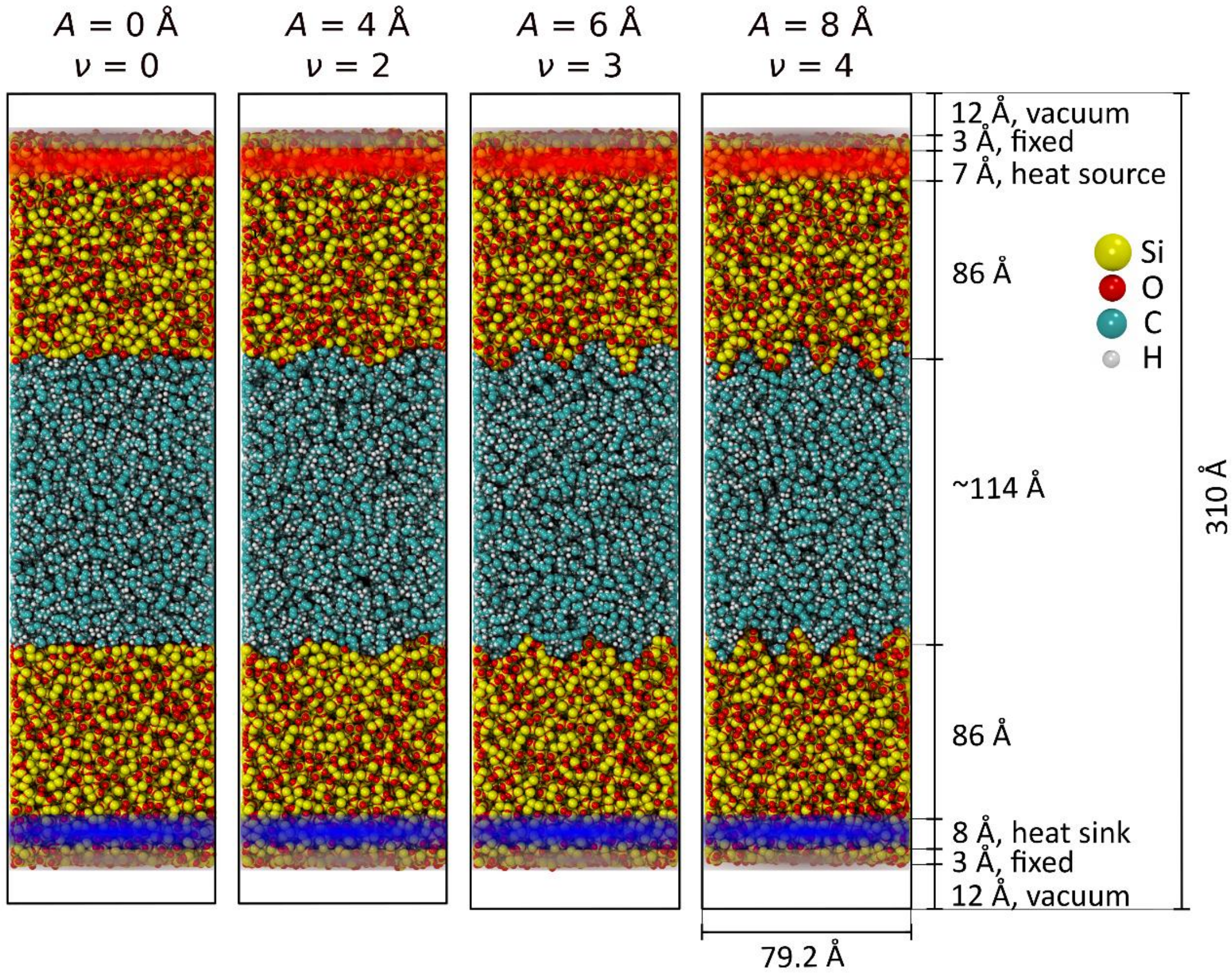

*Figure 1. Representative snapshots of molecular systems with varying amplitudes A and periods ν From left to right: (A = 0, ν = 0), (A = 4, ν = 2), (A = 6, ν = 3) and (A = 8, ν = 4). Heat source (330K) and heat sink (270K) thermostats are shown as regions shaded in red and blue respectively.*

The simulation process started with heating and equilibrating the system at 300 K for 1.5 ns within the NVT ensemble. Subsequently, the topmost and bottommost atomic layers (each 3 Å thick) were anchored along the $z$-direction using a spring-like tethering mechanism. To generate a heat flux, we created heat source and sink regions (each 7 Å thick) by directly rescaling atomic velocities via *temp/rescale* to maintain temperatures of 330 K and 270 K, respectively. The temperature difference imposed in the NEMD simulations was chosen to generate a stable, well-resolved steady-state heat flux while maintaining hexadecane in the fully liquid state. At every rescaling step, the change in total kinetic energy of the atoms in the hot and cold slabs was computed and accumulated. The system then evolved in the NVE ensemble for 2 ns to establish a steady-state heat flow, where the heat produced by the source matched that removed by the sink. Over the following 5 ns, we collected data to compute the average energy flux through the system and obtained two-dimensional temperature and density profiles.

To confirm the consistency of the ITR results, each nanostructured surface configuration was simulated three times, using independently generated initial atomic velocities and varying the vertical offset ($h$) that defines the sinusoidal surface cut, thus sampling different interface morphologies.

The heat flux, $J$, was determined using the formula:

$$J = \frac{\langle E \rangle}{\Sigma \cdot t} \quad (5)$$

where $\langle E \rangle$ represents the average energy transferred across the cross-sectional area $\Sigma$ over time $t$.

To compute the ITR, we first constructed a steady-state temperature profile along the $z$-axis with a spatial resolution of 0.5 Å via binning. The temperature jump, $\Delta T$, at the interface was obtained by extrapolating the linear temperature gradients of the solid and liquid phases to the interface position $z_{eq}$ and calculating their difference.

The interface position, $z_{eq}$, was determined using the Gibbs dividing surface concept [45], which balances the excess density on either side of the interface. Given the discrete density profile from simulations, we approximated $z_{eq}$ by defining a transition layer around the interface, ranging from $z_0$ to $z_0 + d$, where the density shifts from $\rho_1$ (solid) to $\rho_2$ (liquid). We calculated the average density, $\langle \rho \rangle$, within this layer and applied the formula [46] :

$$z_{eq} = z_0 + \frac{\langle \rho \rangle - \rho_2}{\rho_1 - \rho_2} \cdot d \quad (6)$$

Finally, the interfacial thermal resistance, $R_{th}$, was calculated as:

$$R_{th} = \frac{\Delta T}{J} \quad (7)$$

To further investigate the influence of local surface geometry on interfacial thermal transport, we analyzed the thermal resistance separately for regions of positive and negative surface curvature. For each structured surface, we identified the local maxima and minima of the sinusoidal profile, corresponding respectively to convex ("hill") and concave ("valley") regions. The positions of these features were computed analytically based on the sinusoidal parameters:

$$\begin{aligned} x_{hill} &= x_{min} + \frac{a}{4} + k \cdot a, \\ x_{valley} &= x_{min} + \frac{3a}{4} + k \cdot a, \end{aligned} \quad (8)$$

where $k \in [0, \nu - 1]$.

For each identified hill and valley position, we defined narrow sampling stripes of width $\Delta x = a/4$ , extending laterally in both directions from the reference point (see Figure 2a, d and Figure 11). This ensured non-overlapping regions that selectively probed either convex or concave local curvatures at the interface. Due to the periodic nature of the surfaces, each sampling stripe encompassed one interface region where the solid surface exhibited positive curvature, and the opposing surface exhibited negative curvature (or vice versa).

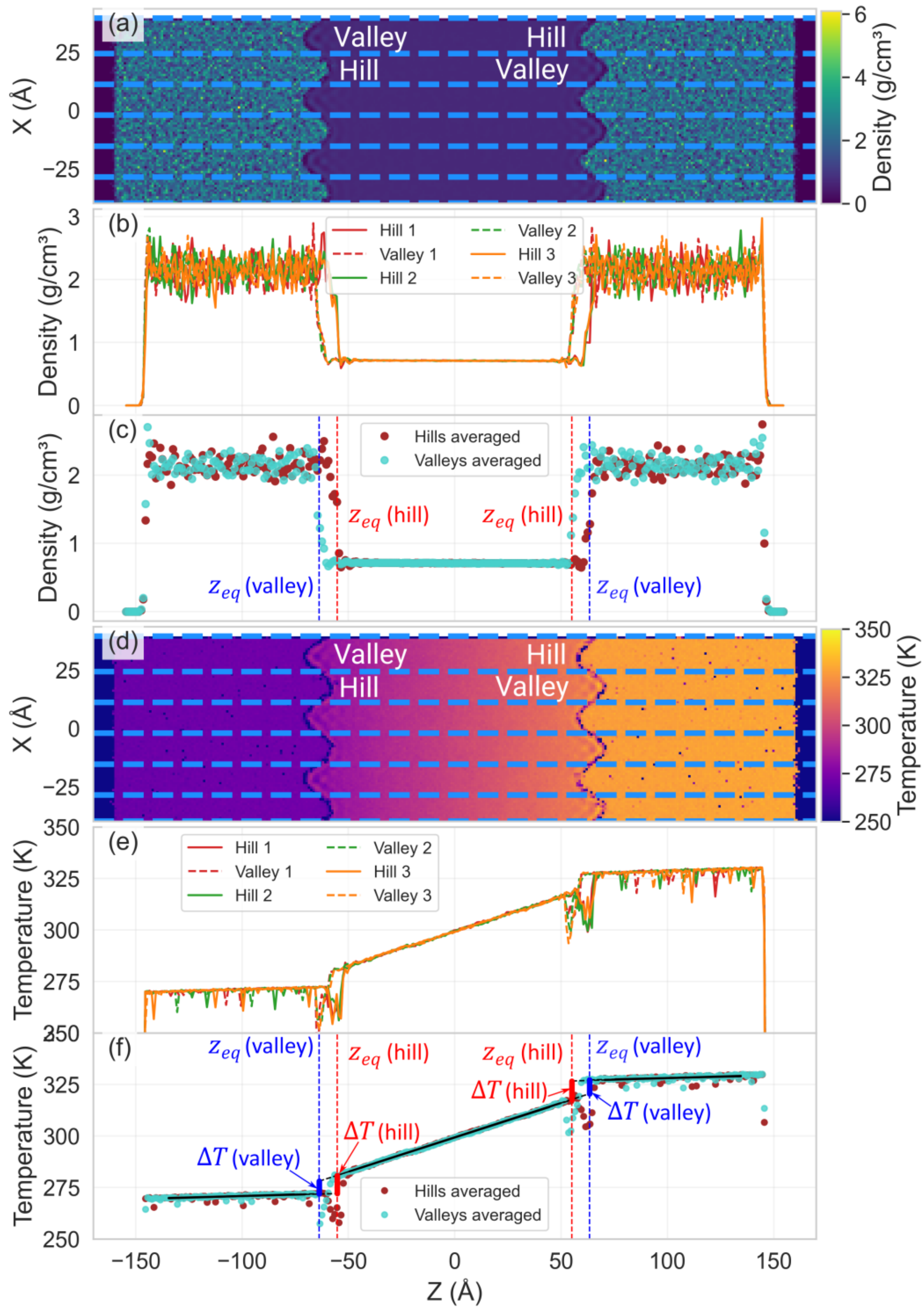

Figure 2. *Dependent on curvature sign $\Delta T$ calculation procedure for the case $A$ = 6 Å, $\nu$ = 3: (a) 2D density distribution. Blue horizontal lines indicate the strips used to compute the 1D profiles; (b) 1D density profiles of each strip; (c) averaged density profiles with $z_{eq}$; (d) 2D temperature distribution; (e) 1D temperature profiles of each strip, (f) averaged temperature profiles with linear fits (black lines) and $\Delta T$.*

We extracted one-dimensional temperature and density profiles from each stripe individually. As a result, each surface morphology yielded two sets of stripe-averaged profiles: one corresponding to regions beginning at convex features and another at concave features, with the reference point always defined on the cooler slab (Figure 2b, e). These profiles were then averaged across all equivalent stripes, producing two representative temperature and density distributions per system — one for "hill" regions (on the cooler side) and one for "valley" regions (Figure 2c, f).

The interfacial position within each stripe was determined independently using the Eq. ( 6 ) applied to the stripe-averaged density profiles of hexadecane and silica. Thus, for each configuration, we identified four interfacial locations where temperature discontinuities could be evaluated: convex and concave regions on both hotter and cooler slabs.

Temperature jumps were measured at these locations via linear extrapolation from the adjacent bulk phases. Knowing the global heat flux and the cross-sectional area of each stripe, we calculated the local interfacial thermal resistance for convex and concave regions separately, using the standard Eq. ( 7 ).

## 2.3. Work of adhesion calculation

The work of adhesion, denoted $W_{adh}$, was evaluated using the "phantom wall" method [47]. This approach utilizes a repulsive potential engaging exclusively with hexadecane molecules and enabling separation of the liquid from the substrate. By progressively shifting the wall and monitoring the resulting forces, the energy needed to disengage the liquid from the surface is determined, directly reflecting the work of adhesion. This method is widely appreciated for its conceptual clarity and compatibility with any interatomic potential [46], [48], [49], [50], [51].

In this study, the system comprised a modulated silica slab entirely overlaid with a hexadecane layer, its thickness established at three times the Lennard-Jones cutoff distance. The simulation domain spanned 79.2 Å along the *x*-axis, 91.4 Å along the *y*-axis, and 85 Å along the *z*-axis. The silica substrate maintained a minimum thickness of 10 Å at its shallowest point along the *z*-direction to exceed the interaction cutoff. A total of 350 hexadecane molecules coated this substrate.

The system was initially stabilized at 300 K over a period of 1.25 nanoseconds. Following this, a repulsive "phantom wall", a virtual plane perpendicular to the *z*-axis, was introduced beneath the interface, positioned beyond the cutoff distance for hexadecane interactions to avoid premature engagement. This wall interacts only with hexadecane molecules (carbon and hydrogen atoms) while remaining imperceptible to the silica substrate. The wall's repulsive force was governed by a 12-6 Lennard-Jones potential, characterized by parameters $\epsilon$ = 6.9382 Å ), $\sigma$ = 3.16 Å, and a cutoff distance of 2.546 Å, ensuring purely repulsive behavior. The exact potential parameters are not pivotal, provided they sustain a repulsive interaction with hexadecane and permit reliable force calculations.

The wall was incrementally elevated in uniform steps of 0.16(6) Å (6 steps per Å), with the maximum displacement varying by system, typically increasing with the amplitude of the surface's sinusoidal modulation. The process ceased when the interaction energy between hexadecane and the substrate dropped below 0.005% of its starting value, a practical threshold given the persistence of long-range electrostatic effects. At each step, the system underwent thermalization for 0.1 ns, after which the force exerted by the wall on the hexadecane and the system's energy were averaged over 0.5 ns. All computations were performed within the NVT ensemble.

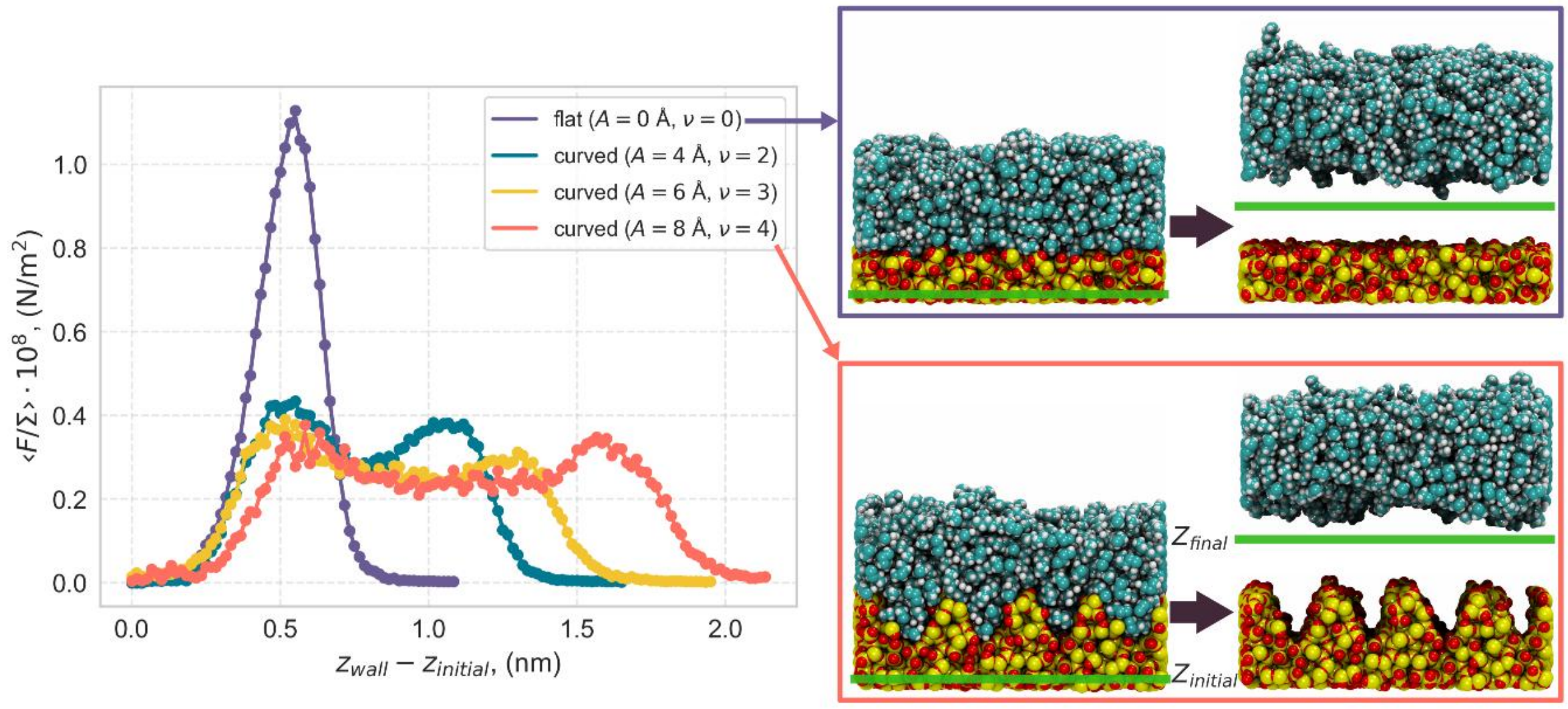

*Figure 3. (a) Scatter plot illustrating the average force per unit area exerted by the phantom wall on hexadecane as a function of wall displacement from its initial position. The data encompass four surface configurations: a flat surface (A = 0 Å, ν = 0), and curved surfaces with increasing amplitude and frequency: (A = 4 Å, ν = 2), (A = 6 Å, ν = 3) and (A = 8 Å, ν = 4); (b) Molecular dynamics snapshots depicting the initial and final states for a flat surface (A = 0 Å, ν = 0), and the most corrugated surface (A = 8 Å, ν = 4). The green line represents the phantom wall.*

We applied thermodynamic integration to calculate $W_{adh}$ from the force-displacement data in Figure 3a. A coupling parameter $\lambda$ (ranging from 0 to 1) was defined to parametrize the wall's quasi-static movement from its initial position $z_{initial}$ to it`s final position $z_{final}$:

$$z_{wall} = \lambda(z_{final} - z_{initial}) + z_{initial} \tag{9}$$

This defines a reversible trajectory within the system's Hamiltonian $H(\lambda)$, bridging the initial ($\lambda$ = 0) and final ($\lambda$ = 1) states. The free energy change is computed via:

$$\Delta F = \int_0^1 \langle {}^{\partial H}/_{\partial \lambda} \rangle d\lambda \tag{10}$$

Here ${}^{\partial H}/_{\partial \lambda}$ represents the force $F(z)$ exerted by the wall on the hexadecane. Converting the variable from $\lambda$ to $z$ and dividing by the cross-sectional area $\Sigma$, the work of adhesion is derived as:

$$W_{adh} = \frac{\int_{z_{initial}}^{z_{final}} F(z)dz}{\Sigma} \quad (11)$$

This integral was numerically evaluated using the discrete force data from Figure 3a.

Knowing the change in potential energy between solid and liquid ($\Delta U$) in the initial and final states, it is possible to estimate the entropy gain per unit area ($T\Delta S$), which is associated with and solid restructuration due to separation [52]:

$$T\Delta S = \Delta U - W_{adh} \quad (12)$$

This relation splits the interfacial free energy change into energy and entropy components [52], [53].

## 2.4. Molecular-scale orientational ordering

To further characterize the molecular-scale structural response of hexadecane to surface curvature, we analyzed the orientational ordering of interfacial molecules in the same simulation setups used for the ITR calculations. For each configuration, 140 statistically independent frames were extracted, separated by 100 000 integration steps, corresponding to a temporal separation of 50 ps. This ensured sufficient decorrelation between sampled configurations.

The orientational order of hexadecane molecules was quantified using the second Legendre polynomial order parameter [54]:

$$P_2 = \frac{\langle 3\cos^2\theta - 1 \rangle}{2}, \quad (13)$$

where $\theta$ is the angle between the molecular axis and the local surface normal and the brackets denote averaging over molecules and time frames. The normal $\boldsymbol{n}(x)$ at the lateral position of the molecular center of mass was obtained from the silica surface profile $z(x)$ (see Eq. ( 1 )) through:

$$\boldsymbol{n}(x) = \frac{\left(-\frac{dz}{dx},\ 0,\ 1\right)}{\sqrt{1 + \left(\frac{dz}{dx}\right)^2}} \quad (14)$$

The molecular axis was defined as the vector connecting the first and last carbon atoms of each hexadecane chain.

To resolve spatial variations, the system was divided into slabs along the surface-normal direction, and $P_2(z)$ was computed by averaging over all molecules whose centers of mass fell within a given slab. For each slab, the mean value and standard error were obtained by averaging over all sampled frames. Statistical uncertainty reflects frame-to-frame fluctuations and was estimated as the standard error of the mean value.

To extract a representative interfacial orientational order parameter for each system, we defined the interfacial region based on $z_{eq}$ determined from the density profiles. Specifically, two interfacial slabs were considered, extending 10 Å into the liquid phase from the upper and lower Gibbs dividing surfaces. Within these regions, an occupancy-weighted average orientational order parameter $P_2^{int}$ was computed, using the number of molecular samples in each slab as weights. The resulting interfacial orientational order parameters were then used for comparative analysis across surfaces with different geometrical characteristics.

## 2.5. Entropy change decomposition

To obtain microscopic insight into the entropic mechanisms governing hexadecane detachment from structured silica surfaces, we performed an explicit decomposition of the entropy change into orientational ($S_{\mathrm{orient}}$), conformational ($S_{\mathrm{conf}}$), and translational ($S_{\mathrm{trans}}$) contributions. All entropy terms were evaluated using Shannon's information entropy applied to appropriately chosen molecular degrees of freedom and were resolved spatially along the surface-normal direction.

For surfaces with nonzero curvature, entropy variations are strongly localized near the solid–liquid interface and cannot be reliably captured by comparing only fully adsorbed and fully detached states. To account for this spatial heterogeneity, entropy contributions associated with translational motion parallel to the interface and molecular orientation were evaluated from the same molecular dynamics trajectories used for the ITR calculations, by resolving local probability distributions in slabs parallel to the interface. In contrast, entropy contributions associated with motion normal to the interface and with molecular conformations were evaluated directly from the phantom-wall detachment trajectories.

### 2.5.1. Spatial binning and definition of interfacial and bulk regions

For each system, the liquid phase was partitioned into bins of thickness $\Delta z = 5$ Å along the surface-normal direction. In every slab, the relevant entropy contribution was evaluated independently, together with the number of corresponding molecular degrees of freedom contributing to that slab. Entropies were computed per degree of freedom and per frame, and subsequently averaged over time.

The interfacial region was defined consistently with the analysis of the orientational order parameter. Specifically, two interfacial zones were identified, extending 10 Å into the liquid phase from the upper and lower Gibbs dividing surfaces $z_{eq}$, as determined from density profiles. All slabs whose centers fell within these regions were classified as interfacial slabs. The bulk entropy was obtained by averaging over slabs located sufficiently far from the interface, where the entropy profiles reach a plateau.

For each entropy contribution $S$, an occupancy-weighted interfacial entropy was computed as

$$S^{\mathrm{int}} = \frac{\sum_{k \in \mathrm{int}} N_k S_k}{\sum_{k \in \mathrm{int}} N_k}, \tag{15}$$

Where $S_k$ is the entropy par degree of freedom in bin $k$, and $N_k$ is the number of corresponding degrees of freedom sampled in that bin. An analogous definition was used for the bulk entropy $S^{\mathrm{bulk}}$.

The entropy change per degree of freedom associated with the interface was then defined as

$$\Delta S = S^{\mathrm{bulk}} - S^{\mathrm{int}}, \tag{16}$$

The extensive entropy change $\Delta S_{total}$ was obtained by multiplying $\Delta S$ by the total number of degrees of freedom contributing to the interfacial region. All entropy changes were converted to interfacial free-energy contributions per unit area as:

$$T\Delta S = \frac{k_B T \, \Delta S_{total}}{\Sigma}, \tag{17}$$

### 2.5.2. Orientational entropy

The orientational entropy quantifies the loss of rotational freedom due to preferential molecular alignment near the surface. Each hexadecane molecule was represented by a single orientational degree of freedom defined by its molecular axis $\boldsymbol{u}$, constructed as the vector connecting the first and last carbon atoms of the chain after unwrapping under periodic boundary conditions in the *xy*-plane. Because the silica substrate is sinusoidally structured, molecular orientations were referenced to the local surface normal rather than a global axis. The local normal was calculated via Eq. ( 14 ). For each chain in each sampled frame, the cosine of the angle between the molecular axis $\boldsymbol{u}$ and the local normal at the chain's center-of-mass position $x_{COM}$ was computed:

$$\cos\theta = u \cdot n(x_{\mathrm{COM}}) \tag{18}$$

Within each slab and each sampled frame, the distribution of $\cos\theta \in [-1,\ 1]$ was accumulated into a histogram with 90 uniformly sized bins, yielding a normalized probability distribution $p_i$. The orientational entropy per hexadecane chain of a given frame is defined as:

$$S_{\mathrm{orient}}^{(\mathrm{per\ chain})} = -k_{\mathrm{B}} \sum_i p_i \ln p_i \tag{19}$$

This procedure was repeated independently for each spatial bin, yielding a spatially resolved orientational entropy profile. The interfacial orientational entropy change was extracted using the occupancy-weighted scheme described above.

### 2.5.3. Translational entropy parallel to the interface

The translational entropy parallel to the interface reflects the lateral delocalization of molecular centers of mass. For each frame, the center of mass of every hexadecane molecule was computed and projected onto the $(x, y)$-plane. Within each spatial bin, a two-dimensional histogram $p(x, y)$ was constructed using a uniform grid covering the entire lateral simulation domain. The histogram was normalized such that $\sum_{i,j} p_{ij} = 1$. The translational entropy per molecule associated with lateral motion was computed as:

$$S_{trans,\,xy}^{(per\ mol)} = -k_B \sum_{i,j} p_{ij}\ ln\, p_{ij} \tag{20}$$

As for the other entropy components, this quantity was evaluated independently in each spatial bin and averaged over time. The interfacial contribution was extracted by comparing the interfacial and bulk slab averages.

### 2.5.4. Translational entropy along the surface normal

In contrast to the other entropy components, the translational entropy associated with motion along the surface normal, $S_{\text{trans, z}}$, was evaluated separately using data from phantom-wall simulations, in which hexadecane was either fully adsorbed or fully detached from the surface. For every surface geometry, the system was equilibrated at the same conditions as in the phantom–wall simulations. Then, two statistically independent 1 ns production intervals were recorded: one with the liquid fully adsorbed and another with the liquid fully detached. Each interval was sampled every 5 ps, yielding 200 frames per state. In these simulations, the density profile along the surface normal remains well defined.

For each state, the distribution of molecular centers of mass was evaluated as a function of the surface-normal coordinate $z_n$, defined as the scalar projection of the molecular center-of-mass position onto the local surface normal $\boldsymbol{n}(x)$. The corresponding probability distribution $p(z_n)$ was obtained by histogramming all molecular samples over multiple frames. The translational entropy along the normal was then computed analogously to Eq. ( 19 ). The entropy change $\Delta S_{\text{trans, z}}$ was obtained by directly comparing the fully detached and fully adsorbed states. Given that the system contains $N_{chains}$ = 350 molecules, the total z-translational entropy change is then:

$$\Delta S_{\text{trans, z}} = N_{chains} \cdot \left( \left\langle S_{\text{trans, z}}^{\text{(per chain)}} \right\rangle_{det} - \left\langle S_{\text{trans, z}}^{\text{(per chain)}} \right\rangle_{\text{ads}} \right) \tag{21}$$

Finally, the translational entropy change along the surface normal, $T\Delta S_{\text{trans, z}}$, is obtained similarly to Eq. ( 17 ).

### 2.5.5. Conformational entropy

To quantify the contribution of intramolecular flexibility to the overall entropy change, we evaluated the conformational entropy of hexadecane molecules through their torsional degrees of freedom. Conformational entropy was evaluated directly from the Phantom Wall detachment trajectories in order to consistently capture the change in internal flexibility of hexadecane chains between the adsorbed and detached states.

For every sampled frame, all hexadecane chains were identified by detecting carbon–carbon connectivity using a distance cutoff of 1.7 Å under periodic boundary conditions in the $x-$ and $y-$directions. Each hexadecane molecule appears as a linear connected component of 16 carbon atoms, which enables deterministic ordering of the chain by following the carbon–carbon adjacency graph from one terminal carbon to the other.

A $C_{16}$ molecule contains. For each chain, all C–C–C–C dihedral angles were computed using the standard geometric definition, resulting in 13 torsional degrees of freedom per $C_{16}$ molecule. For a given frame, the set of dihedral angles $\{\phi\}$ was binned uniformly over the interval $[-180°, 180°]$ using 90 bins, and the conformational entropy per torsion, $S_{\mathrm{conf}}^{(\mathrm{per\ tors})}$, was computed as a Shannon entropy analogously to Eq. ( 19 ), where $p_i$ denotes the probability of observing dihedral angles within bin $i$.

For each system, orientational entropies were averaged over a set of initial (adsorbed) and final (detached) frames. The conformational entropy change upon detachment is then calculated similarly to Eq. ( *21* ):

$$\Delta S_{conf} = N_{tors} \cdot \left( \left\langle S_{conf}^{(\mathrm{per\ tors})} \right\rangle_{det} - \left\langle S_{conf}^{(\mathrm{per\ tors})} \right\rangle_{\mathrm{ads}} \right) \tag{22}$$

Where $N_{tors} = 13 \cdot N_{chains}$ is the number of torsions per frame. Finally, the corresponding entropic contribution was converted to an interfacial quantity similarly to Eq. ( 17 ).

## 2.6. Depletion length calculation

For curved solid–liquid interfaces, conventional one-dimensional density profiles along the Cartesian $z$ direction are insufficient, as the local orientation of the interface varies along the surface. To account for this geometrical complexity, we analyze density variations using a coordinate constructed along the local surface normal. The unit surface normal vector $\boldsymbol{n}(x)$ is obtained directly from the analytical surface profile introduced earlier (Eq. ( 14 )), which was already employed in the analysis of molecular orientational ordering. Using this normal, we define a scalar normal coordinate $s$ for each particle as the projection of its instantaneous position vector $r$ onto the local surface normal,

$$s = \boldsymbol{r} \cdot \boldsymbol{n}(x) \tag{23}$$

In this definition, $s = 0$ lies in the central region of the liquid slab, while the two solid–liquid interfaces are located at finite positive and negative values of $s$. This choice provides a continuous and symmetric mapping of the entire simulation domain and allows both interfaces to be treated simultaneously within a single normal-coordinate framework, without introducing separate reference surfaces.

Two-dimensional, time-averaged mass-density maps were first computed in the $(x, z)$ plane for both hexadecane and silica using the same simulation systems as those employed for the ITR calculations. These maps were then projected onto the normal coordinate $s$ by aggregating mass contributions into bins of $s$ and normalizing by the corresponding bin volumes. This volume-weighted mapping yields one-dimensional density profiles $\rho_{hex}(s)$ and $\rho_{silica}(s)$ that correctly recover bulk densities far from the interface and avoid artefacts associated with simple arithmetic averaging. Bulk reference densities $\rho_{hex}^{bulk}$ and $\rho_{silica}^{bulk}$ were extracted directly from the normal-coordinate density profiles by averaging over well-defined regions deep inside the liquid and solid phases, where the profiles exhibit stable plateaus and are unaffected by interfacial structuring. Figure 4 illustrates an example of normal-coordinate density profiles for the (A = 6 Å, ν = 3) system.

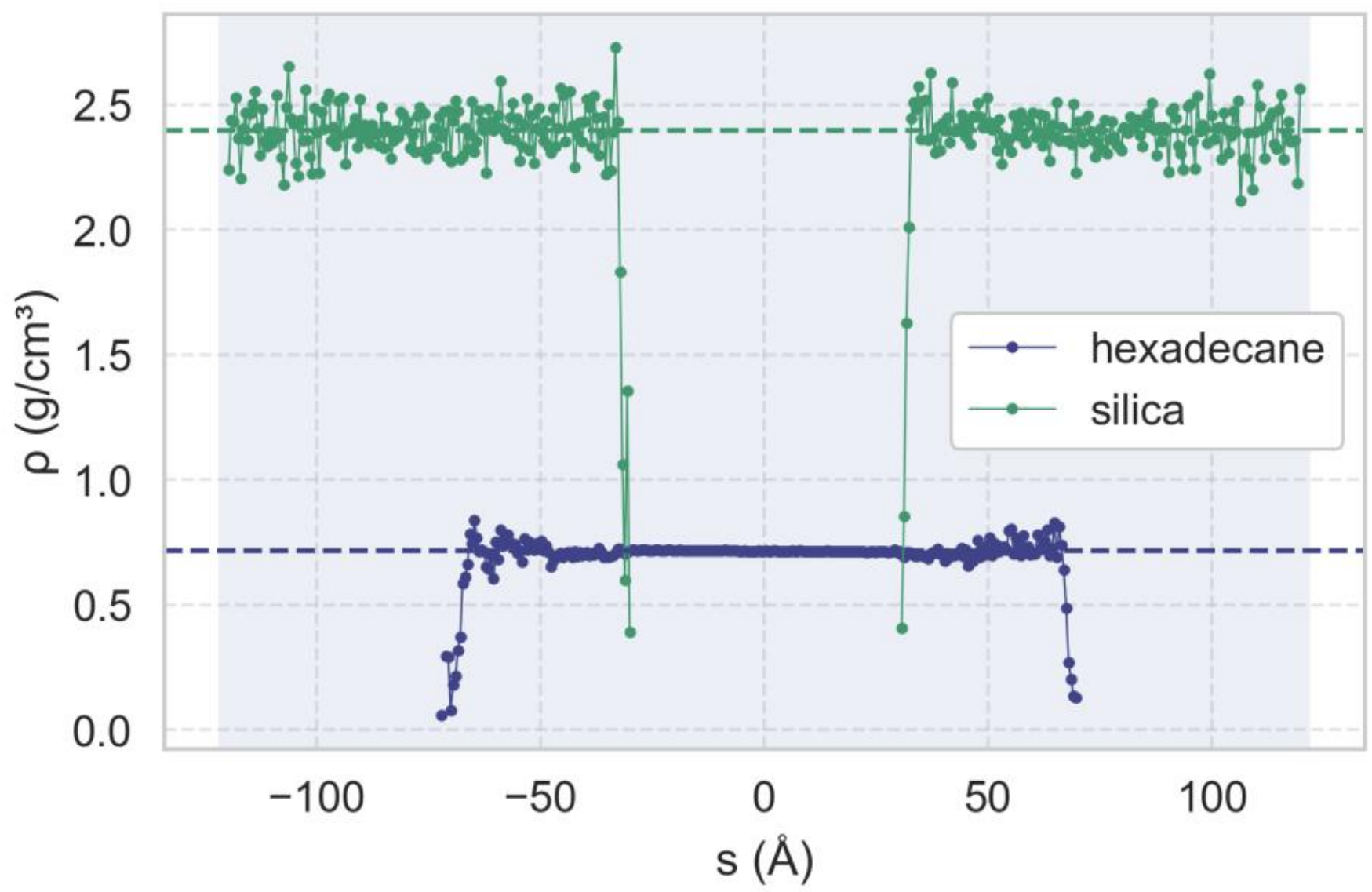

*Figure 4. Normal-coordinate density profiles of hexadecane and silica for the (A = 6 $Å$, $\nu$ = 3) system. Dashed horizontal lines indicate the corresponding bulk densities.*

To quantify the extent of density depletion induced by interfacial curvature, we define a normal-coordinate depletion length, denoted δ, as [55]:

$$\delta = \int_{0}^{\infty} \left[ 1 - \frac{\rho_{silica}(s)}{\rho_{silica}^{bulk}} - \frac{\rho_{hex}(s)}{\rho_{hex}^{bulk}} \right] ds \qquad (24)$$

With this definition, the integrand vanishes in the bulk regions, where one phase attains its bulk density while the other is absent. As a result, only the interfacial regions, where both density profiles deviate from their respective bulk values, contribute to $\delta$. Importantly, the integration is performed over the entire normal-coordinate domain, allowing the depletion length to naturally capture extended density modulations and interference patterns that emerge for strongly curved surfaces. The quantity $\delta$ should therefore be interpreted as an integrated measure of interfacial density deficit, rather than as a geometric thickness associated with a single density layer or a specific structural feature.

## 3. Results and discussion

### 3.1. Spatial density and temperature distributions

The spatial density and temperature distributions in solid and liquid are presented for different structurations of the solid substrate in Figure 5 and Figure 7 respectively. As the representative cases, we chose the following structuration: a flat surface ($A$ = 0 Å, $\nu$ = 0) and three patterned surfaces with increasing amplitudes and numbers of period ($A$ = 4 Å, $\nu$ = 2); ($A$ = 6 Å, $\nu$ = 3); ($A$ = 8 Å, $\nu$ = 4). These four cases span a broad range of surface curvatures, allowing a visual comparison of interfacial density patterns as the interface geometry evolves.

For the 2D density maps and the 1D density profiles, convergence to a constant bulk density occurs far from the interface. For both hexadecane and silica, the density stabilizes at values corresponding to their respective bulk densities 0.719±0.04 g/cm$^3$ and 2.13±0.02 g/cm$^3$.

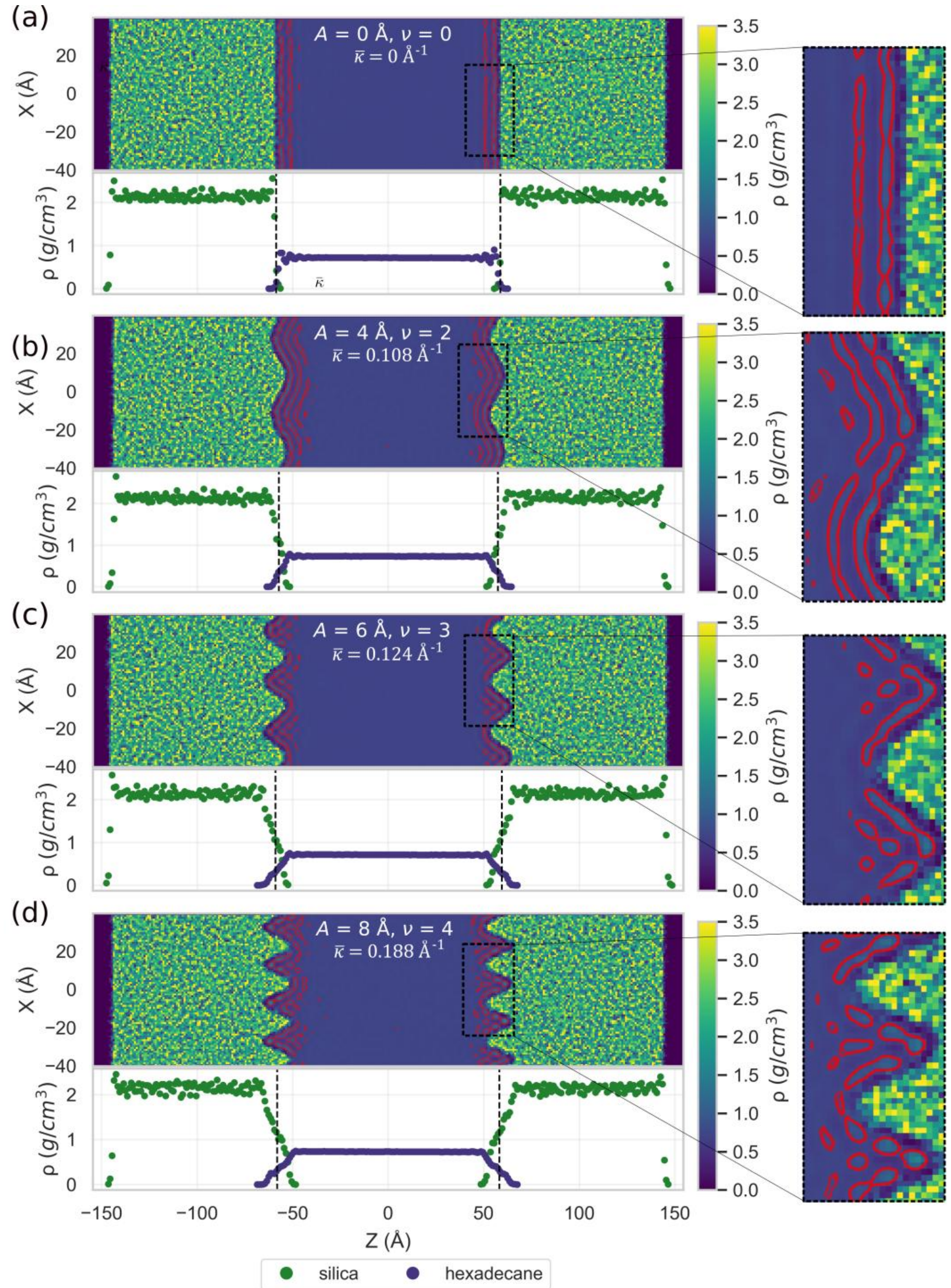

*Figure 5. 2D spatial density maps for four representative cases: (a) (A = 0 Å, $\nu$ = 0); (b) (A = 4 Å, $\nu$ = 2); (c) (A = 6 Å, $\nu$ = 3); (d) (A = 8 Å, $\nu$ = 4). Red contour lines highlight a narrow hexadecane density range near the bulk liquid density, emphasizing the interfacial layering. Below each map, 1D profile of $\rho$ is plotted against z-coordinate, with dashed lines indicating the equilibrium interface $z_{eq}$.*

Close to the interface, the 2D density maps reveal the presence of oscillatory behavior in the liquid density on the solid substrate due to the layering of the hexadecane molecules on the silica, characteristic of the solid-liquid interface [56].

We observed that the spatial pattern of density fluctuations near the solid-liquid interface strongly depends on the interface's curvature. For flat interfaces and those with moderate curvature (Figure 5 a and b), the oscillation of the high- and low-density regions closely follow the geometric contour of the interface. In 2D density maps, the distribution of density maxima and minima essentially mirrors the interface profile, indicating that the surface structure directly templates molecular layering. However, as the interface curvature increases, these fluctuations begin to interfere with each other spatially, producing complex patterns reminiscent of optical interference phenomena. In this regime, the density modulation no longer follows the local geometry alone, but instead reflects the combined influence of neighbouring interface segments, analogous to the overlapping wavefronts in diffraction or interference. This transition suggests a curvature-induced shift from purely surface-templated layering to a regime dominated by collective structural correlations and geometric frustration, characteristic of confined fluids under curved confinement[57].

In addition, we analyzed the 1D density profiles obtained by averaging the 2D fields along the tangential direction. For flat surfaces, the profile exhibits oscillatory behavior near the interface, which correlates with the underlying density layering. In contrast, for curved interfaces, these density oscillations vanish. Instead, we observe a monotonic transition accompanied by a depleted-density region adjacent to the solid surface. Further, for defining the position of the interface, we apply the formal approach based on Gibbs' dividing surface concept [45] with Eq. ( 6 ). The position of this interface ($z_{eq}$) is presented in Figure 5 by dashed lines for each case.

While the two-dimensional density maps and one-dimensional profiles shown in Figure 5 clearly reveal curvature-induced changes in molecular layering, a quantitative metric is required to compare the extent of density depletion across different surface geometries. To this end, we employ the depletion length $\delta$ defined via Eq. ( 24 ), which provides an integrated measure of the density deficit associated with interfacial packing frustration and reflects the cumulative deficit arising from multiple interfacial layers. Figure 6 summarizes the depletion length $\delta$ for all studied surface morphologies.

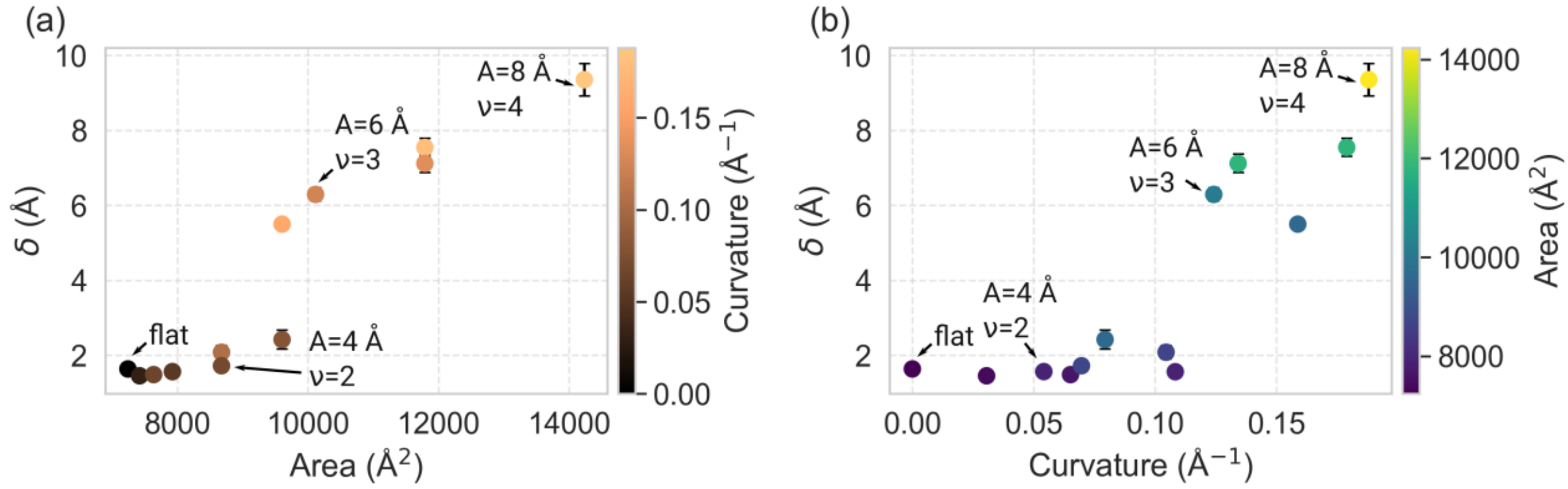

*Figure 6. (a) Depletion length as a function of the effective solid–liquid contact area, with the color scale indicating the mean surface curvature; (b) Depletion length as a function of curvature, with the color scale representing the effective contact area.*

Figure 6a shows that, for comparable effective contact areas, surfaces with higher curvature consistently exhibit slightly larger depletion lengths. However, we observe a pronounced gap in $\delta$ between two systems: (A = 4 Å, ν = 4) and (A = 8 Å, ν = 2), with identical contact areas (≈9600 Å$^2$) but different curvatures (0.158 Å$^{-1}$ against 0.08 Å$^{-1}$ respectively). In this intermediate regime, the depletion length increases abruptly by a factor of approximately 2.25 with curvature, suggesting the onset of a qualitatively different interfacial structuring mechanism.

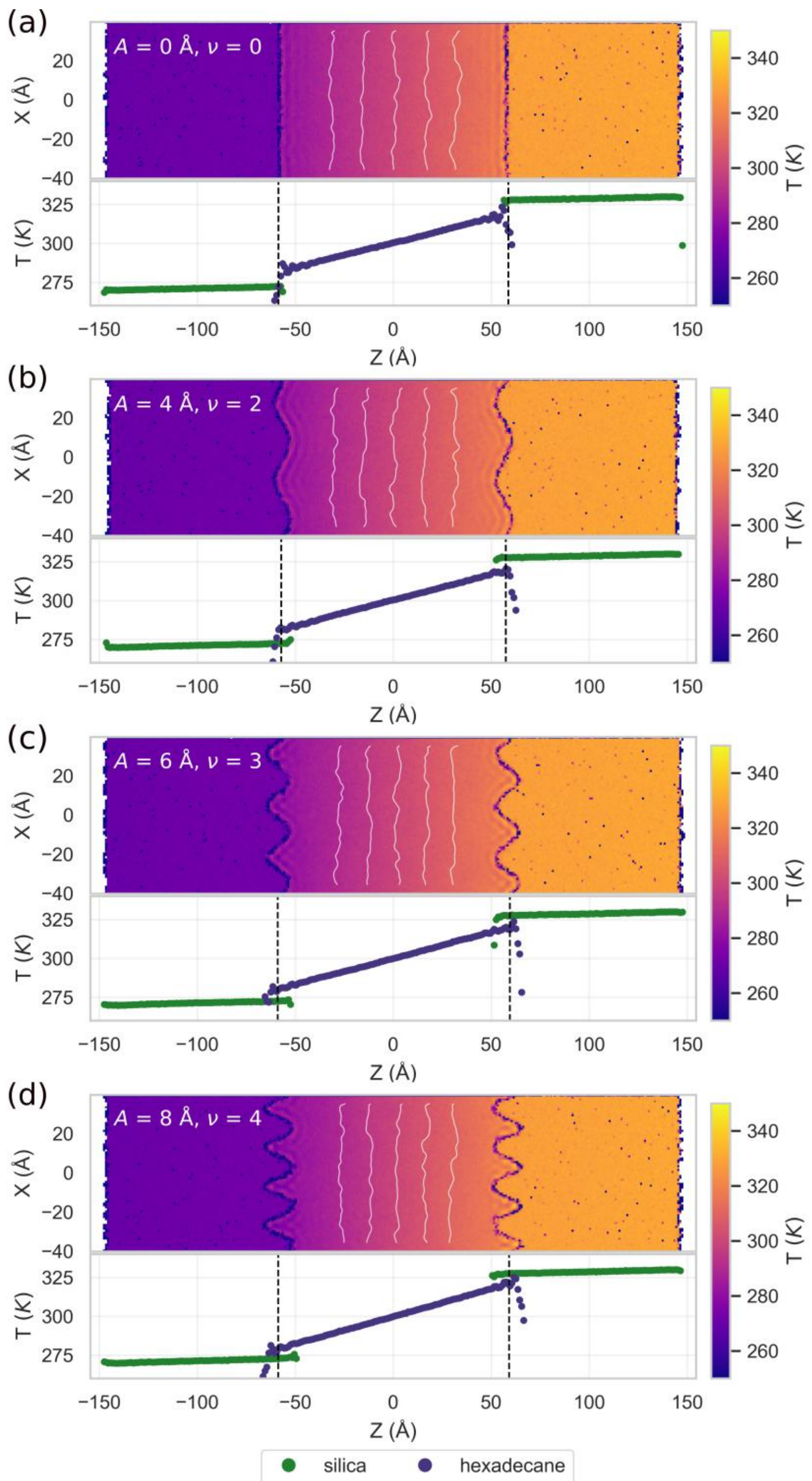

*Figure 7. 2D spatial temperature maps for four representative cases: (a) (A = 0 Å, ν = 0); (b) (A = 4 Å, ν = 2); (c) (A = 6 Å, ν = 3); (d) (A = 8 Å, ν = 4). Below each map, 1D profile of $T$ is plotted against z-coordinate, with dashed lines indicating the equilibrium interface $z_{eq}$.*

A similar nonlinearity is evident when $\delta$ is plotted directly as a function of curvature (Figure 6b). Instead of following a simple linear trend, the depletion length shows a weak dependence on curvature at low values and increases more rapidly beyond a curvature of approximately 0.1 $Å^{-1}$. This nonlinear behavior suggests the presence of a curvature threshold, beyond which interfacial molecular packing becomes increasingly disrupted. This interpretation is consistent with the qualitative features observed in the two-dimensional density maps (Figure 5) and provides further evidence that curvature plays an independent and nontrivial role in controlling interfacial structure.

The 2D temperature distribution maps presented in Figure 7 show that, in the region far from the interface, the isotherms, highlighted by white contour lines corresponding to temperatures between 290 and 310 K with a spacing of 5 K, are oriented parallel to the solid/liquid interface, indicating spatial uniformity in the tangential direction. The contour lines were extracted from a mildly smoothed temperature field (Gaussian smoothing with $\sigma = 1.5$) to improve visual clarity. As shown by the 1D temperature profiles, the temperature varies linearly in both the solid and liquid regions away from the interface, which is characteristic of diffusive thermal transport. This linear gradient enables the estimation of the thermal conductivity of each phase using Fourier's law. The resulting values are $2.8 \pm 0.17$ $W \cdot m^{-1} \cdot K^{-1}$ for silica and $0.21 \pm 0.02$ $W \cdot m^{-1} \cdot K^{-1}$ for hexadecane.

In contrast, close to the interface, the temperature distribution exhibits features that closely correlate with the local density fluctuations. These interfacial temperature modulations reflect the influence of structural layering and interfacial ordering on energy transport at the molecular scale.

## 3.2. Thermal characteristics

A distinct temperature jump is observed at the solid–liquid interface, which is characteristic of interfacial thermal resistance. The magnitude of this jump, defined at the interface using the Gibbs dividing surface concept, is shown in Figure 8a. The temperature jump, calculated in this formal manner, decreases with increasing interfacial area, consistent with improved thermal contact resulting from the higher density of contacting surface regions. However, a clear dependence on interface curvature is also observed: for the same effective contact area as the curvature increases, the temperature jump tends to increase. This behavior may be attributed to packing frustration and curvature-induced suppression of density layering, as discussed above.

To further quantify the efficiency of interfacial thermal transport, the heat flux across the solid–liquid interface was computed and is presented in Figure 8b. As expected, an increase in the effective interfacial area leads to a general trend of enhanced heat flux, reflecting the improved overall thermal transport efficiency in systems with extended or structured interfaces. Nevertheless, an apparent decrease in heat flux for systems with identical effective contact area is observed as the curvature increases. Interestingly, the data reveal the competing influence of effective contact area and surface curvature on heat flux and appear to separate into two distinct trends: one corresponding to lower-curvature surfaces, which yield higher absolute heat fluxes, and another corresponding to higher-curvature surfaces, which result in reduced flux at the same contact area.

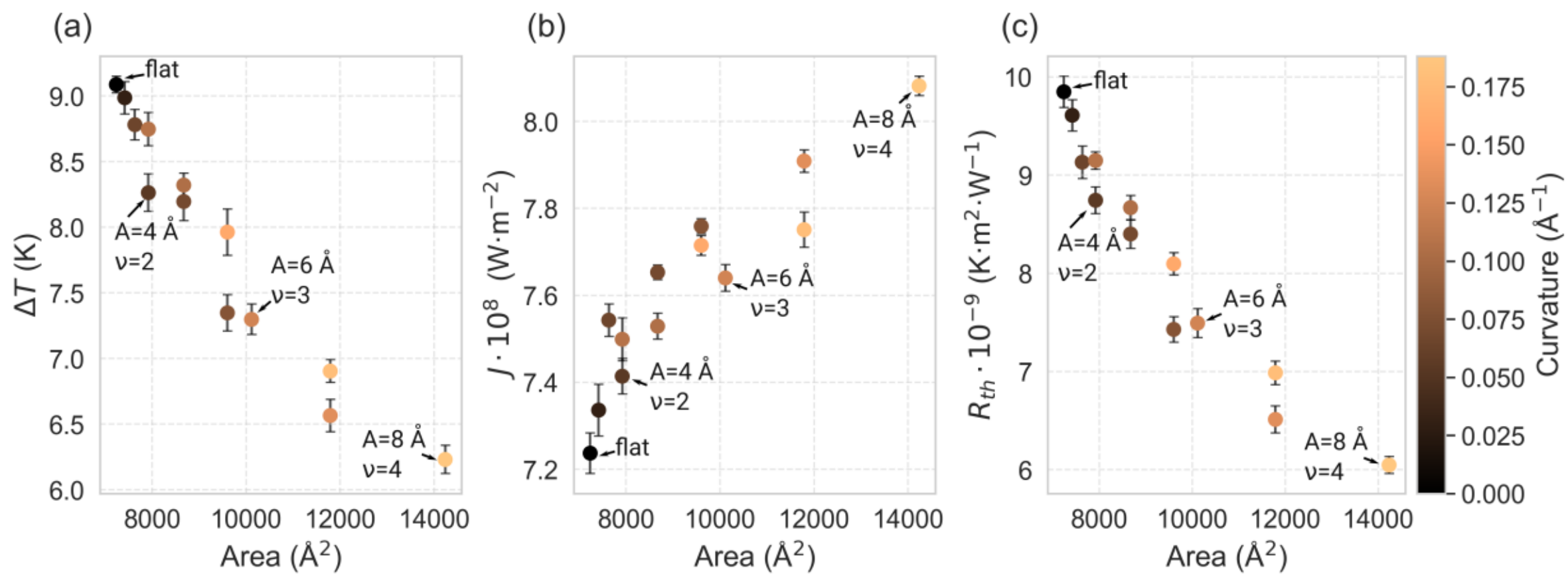

*Figure 8. (a) Temperature jump at the solid–liquid interface, (b) imposed heat flux across the system, and (c) ITR, plotted as functions of the effective contact area.*

Consequently, the ITR calculated from the temperature jump and heat flux decreases (see Figure 6c) as the contact area increases. It should be noted that the absolute magnitude of the interfacial thermal resistance ($6\text{-}10 \cdot 10^{-9}$ $K \cdot m^2 \cdot W^{-1}$) obtained in this work is comparable to experimentally measured values reported for solid‑liquid interfaces, such as the gold‑water interface [58]. Moreover, for configurations with the same contact area, lower-curvature surfaces consistently exhibit lower ITR values. These findings indicate that enhancing thermal transport across solid–liquid interfaces requires both an increased contact area and a reduced surface curvature.

## 3.3. Finite-size effects

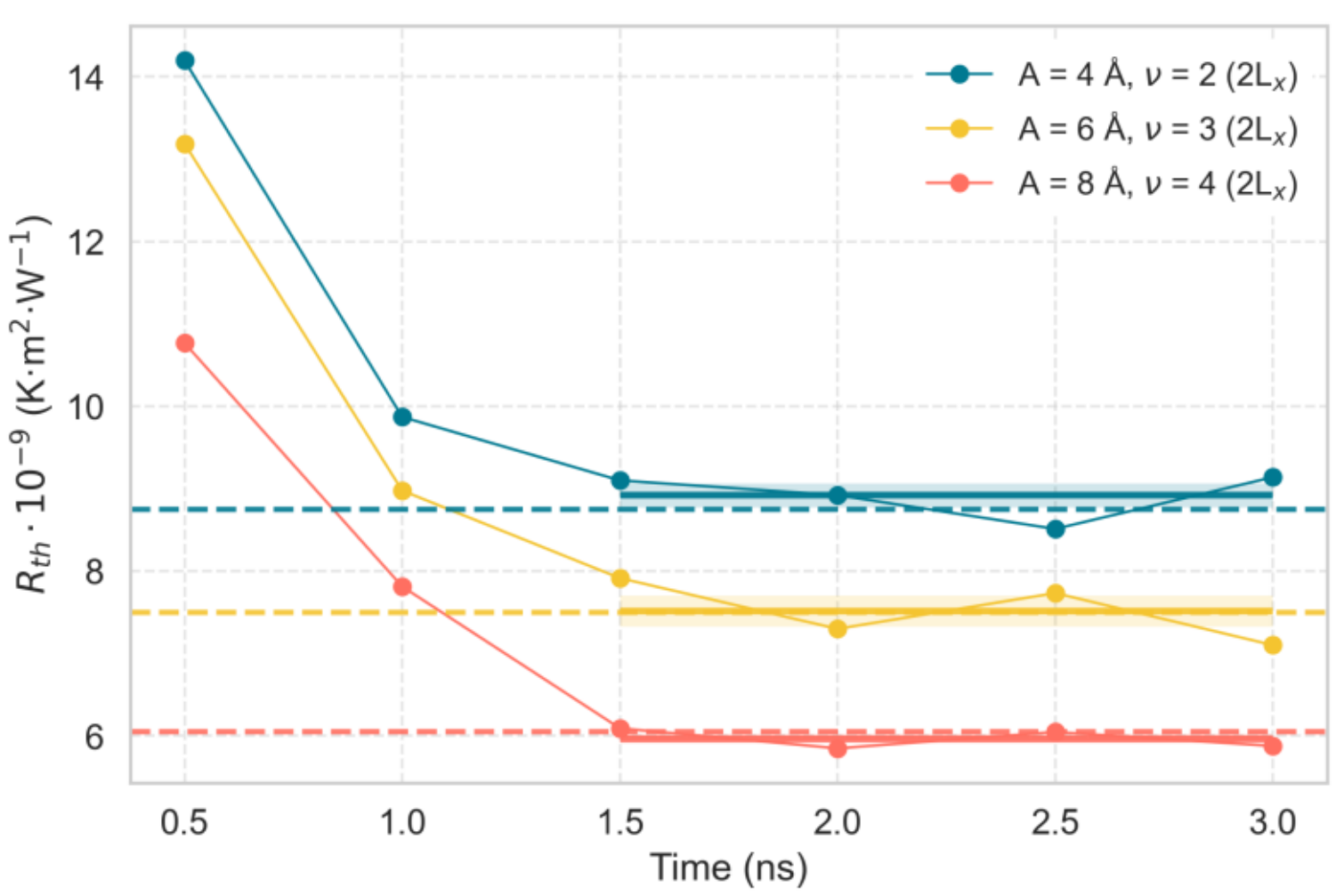

*Figure 9. Time evolution of the ITR for three representative corrugated surfaces: (A = 4 Å, ν = 2), (A = 6 Å, ν = 3) and (A = 8 Å, ν = 4) simulated in laterally enlarged systems ($2L_x$). Solid lines with markers show the instantaneous values, while horizontal solid bands indicate the mean value averaged over the last four data points, with shaded regions representing the associated statistical uncertainty. Dashed horizontal lines correspond to the reference $R_{th}$ values obtained for the original system sizes.*

Since the interfacial modulation is imposed along the transverse direction, one may question whether the calculated interfacial thermal resistance is affected by the lateral system size. To assess possible finite-size effects, we performed additional simulations for three representative surface morphologies: (A = 4 Å, ν = 2), (A = 6 Å, ν = 3) and (A = 8 Å, ν = 4), in which the simulation box length along the x-direction was doubled, while keeping all other parameters unchanged. For these enlarged systems, the same NEMD protocol was applied, and the simulations were extended up to 3 ns of production time following 1.5 ns of thermalization to ensure full convergence of the interfacial thermal resistance. The resulting time evolution of ITR is shown in Figure 9.

In all cases, the ITR converges to stable plateau values that are statistically indistinguishable from those obtained for the original system sizes. The differences between the two system sizes remain well within the estimated statistical uncertainties. These results demonstrate that the lateral dimensions employed in the main simulations are sufficient to capture the interfacial heat transport, and that finite-size effects associated with the imposed transverse modulation are negligible for the system sizes considered in this work.

## 3.4. Molecular-scale orientational ordering

To elucidate the molecular origin of the curvature-dependent interfacial thermal resistance, we analyze the orientational ordering of hexadecane molecules in the vicinity of the solid–liquid interface. Figure 10 shows the interfacial orientational order parameter $P_2^{int}$ as a function of the mean surface curvature for all studied geometries.

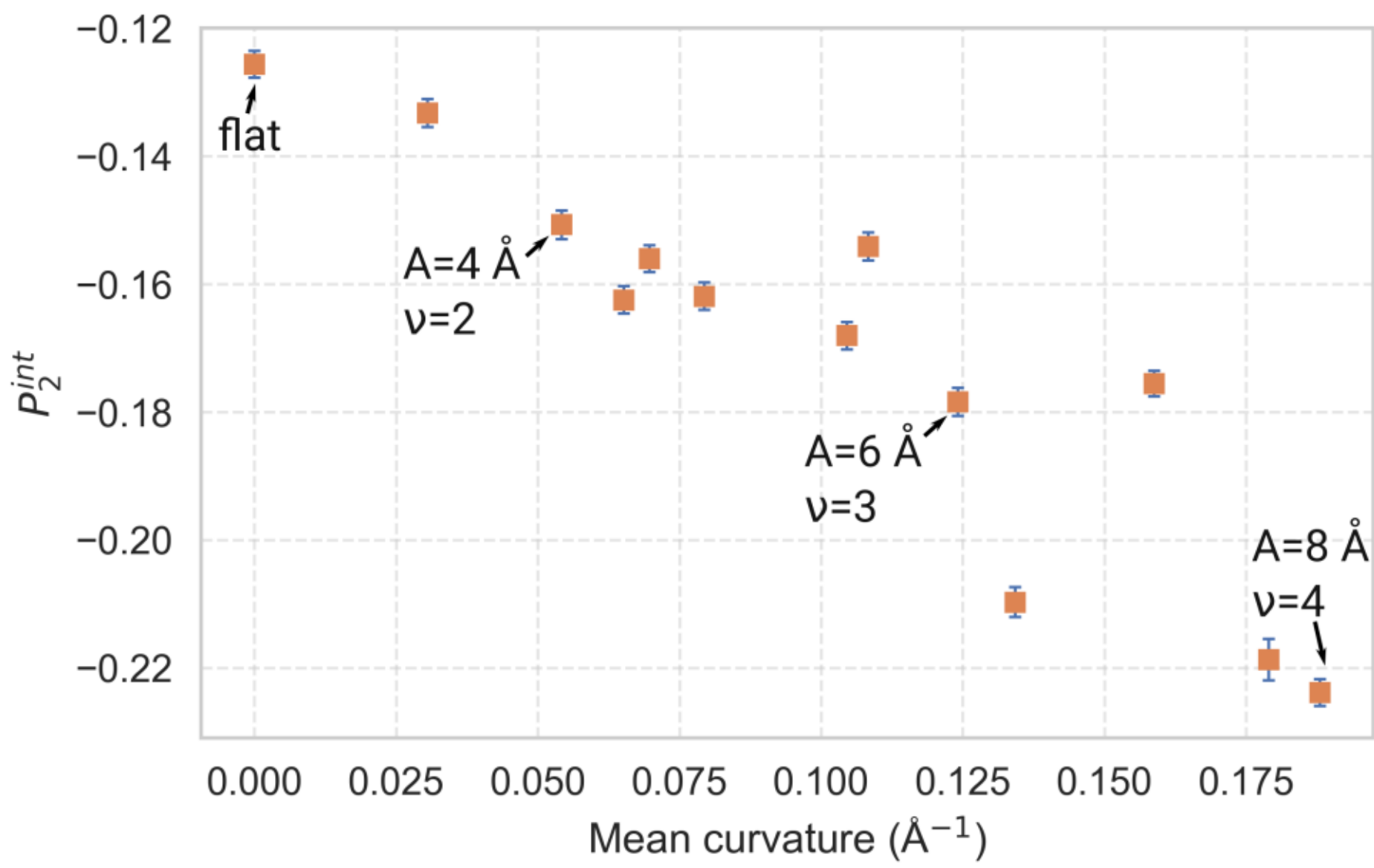

*Figure 10. Interfacial orientational order parameter $P_2^{int}$ of hexadecane molecules as a function of the mean surface curvature.*

For the flat surface, hexadecane molecules exhibit a moderate orientational ordering with respect to the surface normal, consistent with surface-templated layering observed in the density profiles (Figure 5a). As the surface curvature increases, the interfacial orientational order parameter becomes progressively more negative, indicating a systematic reorientation of hexadecane chains toward configurations parallel to the local interface. This trend reflects curvature-induced packing frustration: at low curvature, molecules can partially align with the surface normal, while strongly curved interfaces impose rapidly varying local normals that suppress collective normal alignment. As a result, hexadecane chains preferentially adopt orientations parallel to the interface, minimizing steric frustration but simultaneously disrupting normal-aligned layering. These curvature-induced changes in molecular orientation provide a direct microscopic link between surface geometry and the degradation of interfacial heat transport efficiency.

For all systems, the orientational order parameter evaluated in the bulk liquid region converges to 0 within statistical uncertainty, indicating an isotropic orientational distribution far from the interface. This confirms that the observed variations of $P_2^{int}$ are genuinely interfacial in origin and not an artifact of the analysis procedure.

## 3.5. Local curvature effects on ITR

Following the global ITR analysis, we further decomposed the total interfacial thermal resistance into contributions from convex ("Hill") and concave ("Valley") regions, as defined in the Methodology section. This spatial decomposition allows us to isolate the effect of local curvature sign on thermal transport.

Figure 11 presents the ITR for “hill” and “valley” regions separately, plotted against the effective contact area. Hill regions correspond to positively curved surface segments, whereas valley regions represent negatively curved segments.

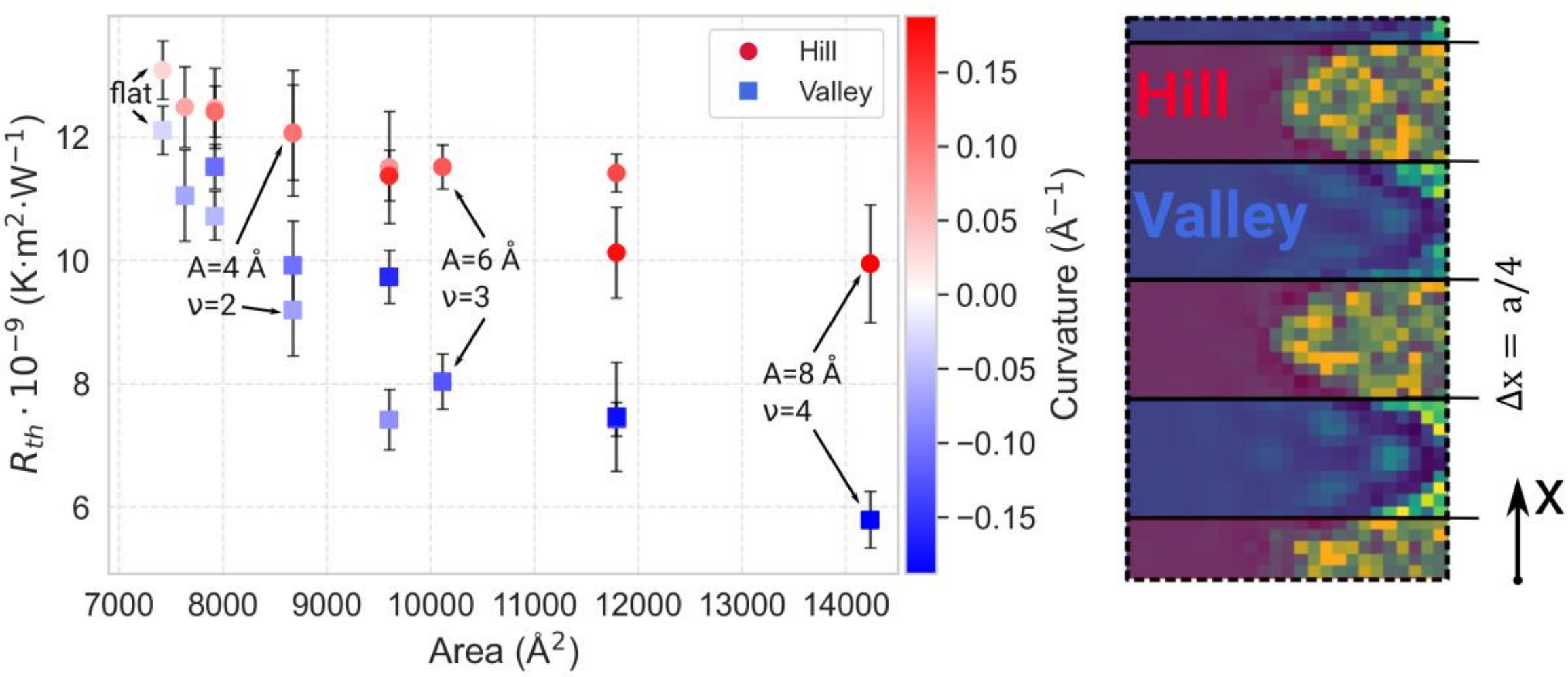

*Figure 11. Interfacial thermal resistance separated into contributions from regions of negative curvature (“valley”, blue squares) and positive curvature (“hill”, red circles), plotted as a function of the effective contact area. Valley regions correspond to concave surface segments shaded in blue on the right, while hill regions correspond to convex segments shaded in red.*

The data clearly reveal a systematic difference in thermal resistance between these two types of regions: valley regions consistently exhibit lower resistance than hill regions across all systems. Moreover, this disparity increases with the effective contact area, and therefore with the absolute curvature of the surface. In the most extreme case ($A$ = 8 Å, $\nu$ = 4), the thermal resistance in hill regions is 1.7 times greater than that in valley regions.

Interestingly, the curvature dependence within each region type exhibits contrasting behavior: for hill regions, increasing curvature (at a constant contact area) decreases thermal resistance. For valley regions, the same increase in curvature generally improves the resistance. This asymmetric trend is non-trivial and suggests that curvature influences thermal transport not only through geometrical confinement but also via molecular-level structural organization represented by the density fluctuations and stress distributions at the interface.

While the present simulations are based on periodic surface morphologies, the decomposition of interfacial thermal resistance into convex ( “hill” ) and concave ( “valley” ) contributions provides a natural bridge to random and fractal-like geometries. Any complex surface can be locally decomposed into regions of positive and negative curvature across multiple length scales. At the molecular level, heat transfer is governed by this local interfacial geometry, such that the effective interfacial thermal resistance can be interpreted as a weighted average of curvature-dependent local resistances. The systematic difference observed between hill and valley regions therefore represents a transferable, local mechanism that is expected to persist in disordered and multiscale interfaces.

asymmetric trend is non-trivial and indicates that curvature influences thermal transport not only through geometrical confinement but also via molecular-level structural organization, reflected in interfacial density modulations and stress distributions.

These results highlight that interfacial thermal resistance is governed by local energetic and structural properties of the interface. Since both adhesion strength and entropy production directly stem from molecular packing, ordering, and confinement at the solid–liquid boundary, a consistent interpretation of the thermal transport behavior requires explicit consideration of the underlying energetic balance associated with interfacial detachment.

In the following section, we therefore analyze the work of adhesion and the associated entropic contributions, and examine how these quantities depend on surface geometry and relate to the observed trends in interfacial thermal resistance.

### 3.6. Work of adhesion and entropic effects

Figure 12 summarizes the interplay between adhesion, entropy, and surface geometry. Panel (a) shows the work of adhesion, $W_{adh}$, the entropy-related energy gain obtained from the phantom-wall method, $T\Delta S_{PW}$, and the total molecular entropy gain, $\sum_i T\Delta S_i$, plotted as functions of the effective solid–liquid contact area. It is important to note that the work of adhesion values are normalized by the projected cross-sectional area. At the same time, the *x*-axis represents the effective contact area, which increases with surface corrugation.

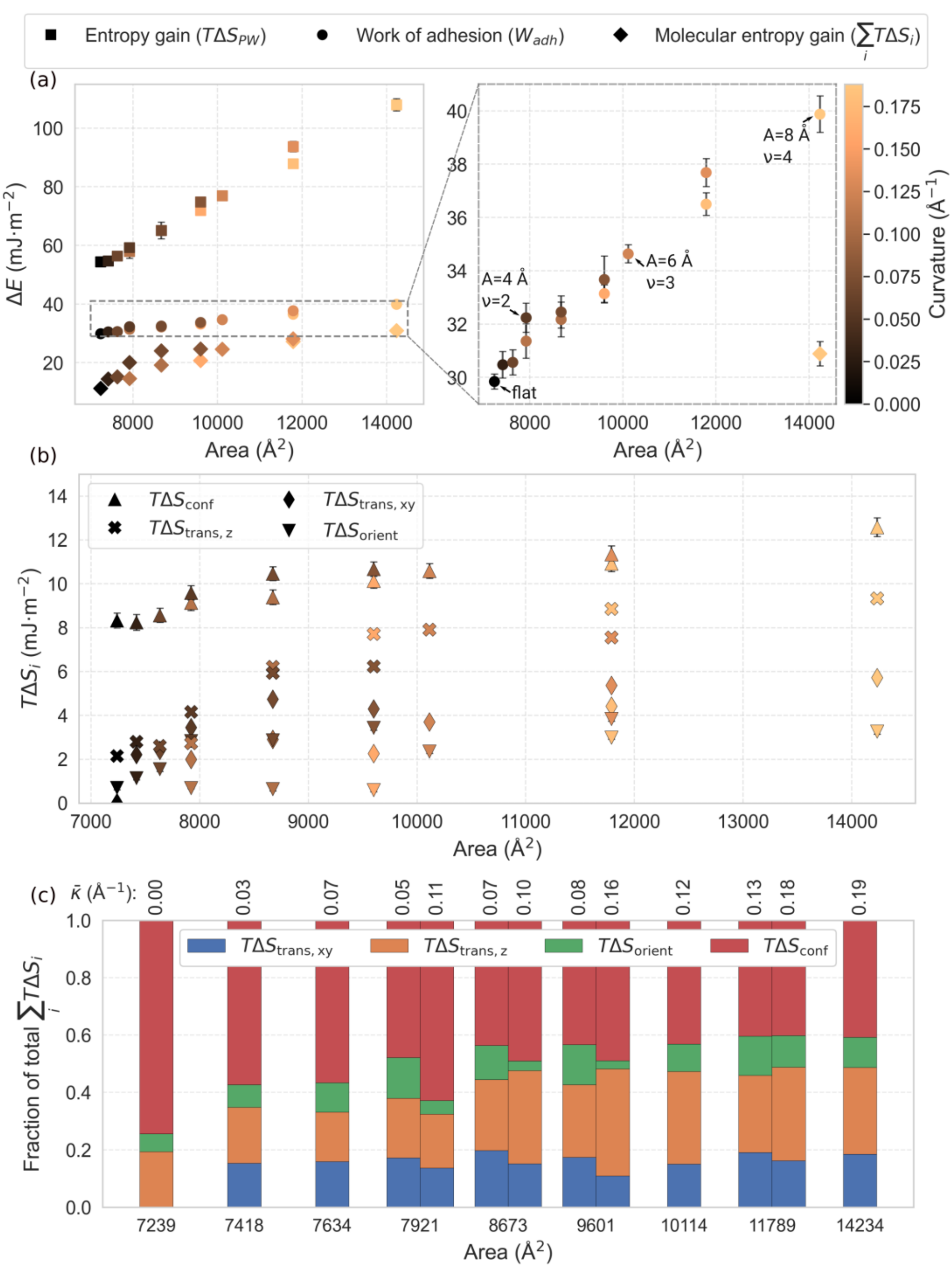

*Figure 12. (a) Work of adhesion (circles), entropy gain obtained from the phantom-wall method (squares), and the total molecular entropy gain (diamonds) as functions of the effective contact area. A magnified view highlights differences in $W_{adh}$ at small curvatures; (b) Decomposition of the molecular entropy gain into translational (normal and lateral), orientational, and conformational contributions; (c) Normalized contributions of each entropy component to the total molecular entropy gain, shown as stacked bars for each system. For identical contact areas, bars are shown side by side, corresponding to surfaces with lower (left) and higher (right) curvature.*

The work of adhesion increases approximately linearly with the effective contact area, which is expected since stronger interfacial contact facilitates greater adhesive interactions. For cases with identical contact area, systems with lower surface curvature exhibit higher work of adhesion, indicating more efficient molecular packing and binding.

In contrast, the entropy-related contribution to the potential energy change obtained from the phantom wall method grows more steeply with surface curvature and exceeds the magnitude of the work of adhesion in all cases. Even for the flat surface ($A$ = 0 Å, ν = 0), this contribution remains substantial, surpassing the work of adhesion across all cases. Intriguingly, at identical contact areas, surfaces with higher curvature display a reduced entropy term compared to those with lower curvature, which also reveals the impact of the reordering of the liquid molecules close to the interface. When comparing this result to our previous study on functionalized but flat silica surfaces [46], we observe that the magnitude of the entropy-related term was comparable to the work of adhesion, whereas in the present case, the structural reorganization dominates the total energy change.

To elucidate the microscopic origins of this entropy gain, panel (b) of Figure 12 decomposes the molecular entropy into conformational ($S_{conf}$), translational (normal ($S_{trans,z}$) and lateral ($S_{trans,xy}$)), and orientational ($S_{orient}$) contributions. The total molecular entropy gain amounts to 10–30 mJ·m$^{-2}$, significantly smaller than the entropy inferred from the phantom-wall method, yet reproducing the same qualitative dependence on surface geometry. Among the molecular contributions, the conformational entropy of hexadecane constitutes the largest single component, followed by translational entropy normal to the interface, lateral translational entropy, and orientational entropy. This hierarchy reflects the partial release of torsional constraints upon detachment, combined with a progressive delocalization of the alkane layer away from the surface.

Panel (c) of Figure 12 highlights how the relative contributions of these entropy components evolve with curvature and contact area. For surfaces with identical contact area, increasing curvature leads to a larger fraction of conformational and normal translational entropy, while the contributions from orientational and lateral translational entropy decrease. This indicates that strong surface corrugation primarily affects chain conformations and confinement normal to the interface, while simultaneously suppressing in-plane mobility and orientational ordering.
Conversely, increasing the contact area reduces the relative contribution of conformational entropy and enhances the role of translational degrees of freedom, consistent with the formation of thicker interfacial layers. Interestingly, the relative contribution of orientational entropy is governed primarily by surface curvature rather than by the contact area. As expected, the lateral translational entropy contribution is negligible for the flat surface.

Importantly, while the molecular entropy decomposition captures the correct trends with curvature and contact area, its absolute magnitude remains significantly smaller than $T\Delta S_{PW}$.
This discrepancy arises because the phantom-wall entropy includes collective and interfacial contributions beyond single-molecule degrees of freedom, such as many-body correlations, interfacial density fluctuations, and energy fluctuations of solid–liquid interactions, which are not accessible through Shannon entropies of marginal distributions. Similar limitations of configurational entropy decompositions have been discussed in previous theoretical studies [59], where neglect of correlated motions leads to systematic underestimation of total entropy changes. The observed reduction of entropy gain with increasing curvature is consistent with earlier studies on alkane monolayers and polymer-like systems, where geometric frustration disrupts collective ordering and partially preserves configurational entropy in the adsorbed state [60], [61], [62], [63], [64].

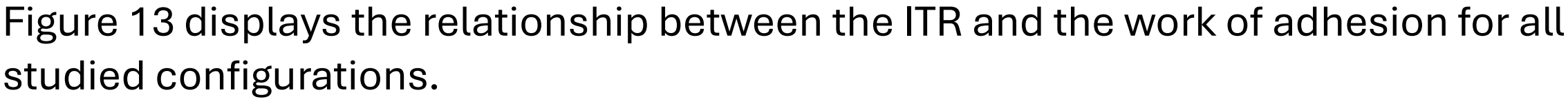
Figure 13 displays the relationship between the ITR and the work of adhesion for all studied configurations.

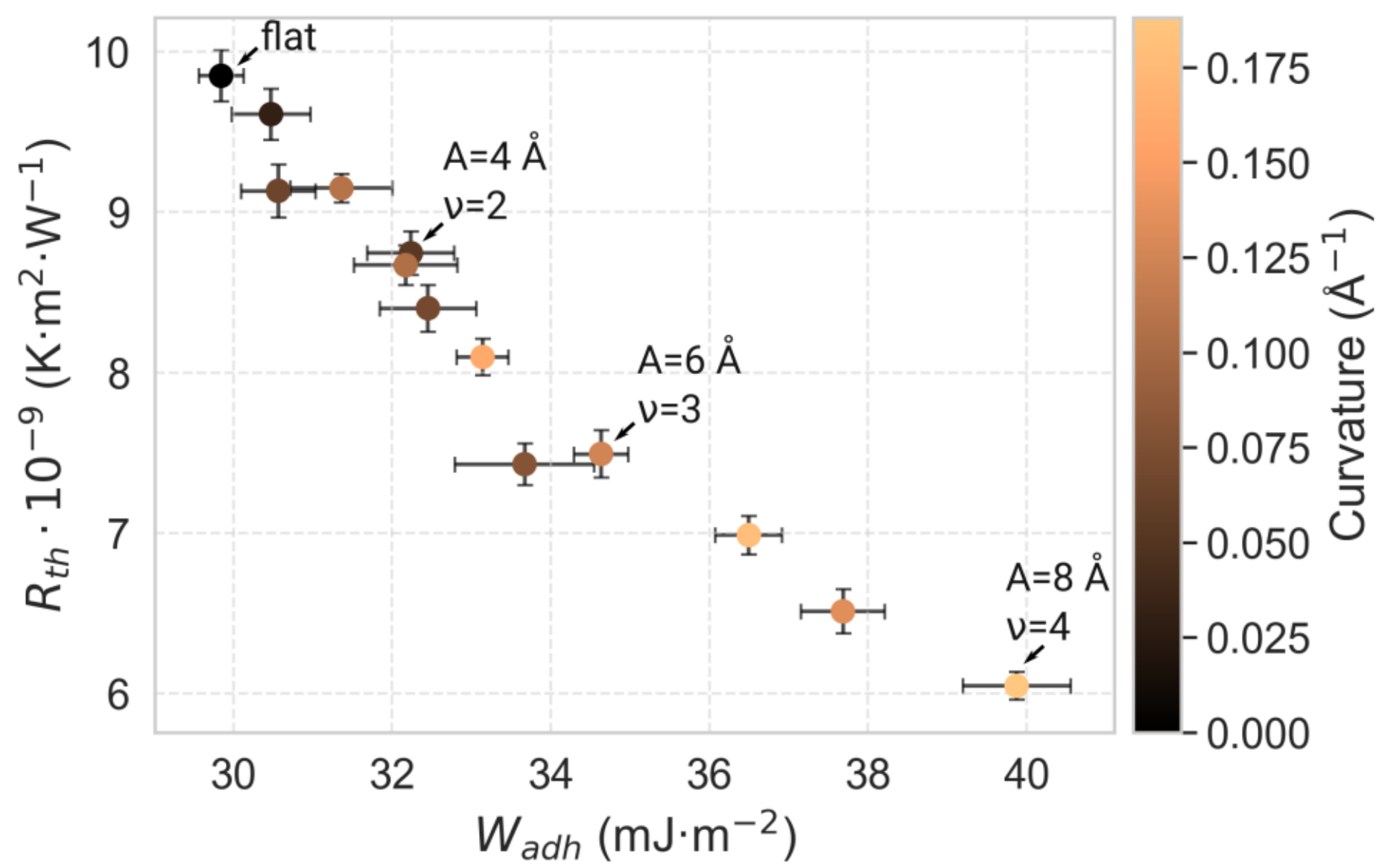

*Figure 13. ITR as a function of the work of adhesion for all studied surface geometries*

The data reveal a general inverse correlation between ITR and the work of adhesion, consistent with commonly reported trends [55], [65], [66], [67], [68]. Initially, ITR decreases nearly linearly with increasing adhesion, indicating enhanced thermal coupling at the interface. However, at high adhesion values, a deviation from linearity emerges. These points correspond to highly curved surfaces, suggesting that beyond a certain level of surface modulation, additional factors, such as interfacial structural ordering or molecular confinement effects, may begin to influence heat transfer significantly. In particular, the near-surface organization of hexadecane molecules could provide additional resistance to energy transport.

## Conclusions

This study demonstrates that surface curvature acts as a fundamental thermodynamic control parameter at solid–liquid interfaces, rather than a purely geometric descriptor. Through systematic molecular simulations of structured silica substrates in contact with hexadecane, we demonstrate that increasing curvature disrupts interfacial molecular ordering, enhances confinement frustration, and fundamentally alters both energy dissipation and heat transport pathways. This disruption is quantitatively captured by a reduction of the orientational order parameter near the interface and increase of depletion length, indicating a progressive loss of molecular normal alignment as curvature increases.

While increased contact area generally promotes stronger adhesion and higher thermal conductance, we show that this trend is non-monotonic: beyond a critical curvature, interfacial thermal resistance increases due to loss of packing efficiency. Our decomposition of the interface into convex and concave regions reveals a pronounced asymmetry in thermal performance, with valley-like regions consistently outperforming hills in heat conduction. This spatial heterogeneity provides a robust framework for understanding and engineering interfacial heat flow.

Energetically, the detachment process is governed not only by adhesion but also by a dominant entropic contribution, which scales nonlinearly with curvature. The phantom-wall entropy captures many-body effects that are inaccessible to Shannon entropies of marginal molecular distributions, including correlated molecular rearrangements, interfacial density fluctuations, and fluctuations of solid–liquid interaction energies. Our results therefore highlight an intrinsic limitation of molecular-level entropy decompositions: although they reliably reproduce qualitative trends with curvature and contact area, they systematically underestimate the total entropic contribution to interfacial free-energy changes. At the molecular scale, we find that the dominant entropy contribution arises from the conformational degrees of freedom of hexadecane, followed by translational entropy along the surface normal. Lateral translational and orientational entropy play secondary roles and are particularly sensitive to surface curvature. Increasing curvature shifts the entropy balance toward conformational recovery and normal delocalization, while suppressing in-plane mobility and orientational ordering.

It is important also to note, that even for complex interfacial geometries, a Gibbs dividing surface provides a consistent definition of the interface, enabling the description of heat transfer through an effective interfacial thermal resistance. From a practical point of view, this representation is advantageous, as it allows interfacial effects to be incorporated directly into continuum-scale thermal models used for PCM system design.

Altogether, our findings establish a direct link between surface geometry, molecular thermodynamics, and interfacial transport. This insight opens new avenues for the rational design of nanostructured materials and thermal interface layers, where geometry can be harnessed to modulate adhesion, control energy dissipation, and enhance heat transfer at the molecular scale.

## Acknowledgements

This research is supported by ANR project “PROMENADE” No. ANR-23-CE50-0008 and by the ANR French PIA project “Lorraine Université d’Excellence” No. ANR-15-IDEX04LUE. Molecular simulations were conducted using HPC resources from GENCI-TGCC and GENCI-IDRIS (eDARI projects No. A0170913052 and A0190913052), as well as resources provided by the EXPLOR Center hosted by the University of Lorraine.

During the preparation of this work, the authors used ChatGPT (OpenAI) to assist with language editing, text refinement, and clarification of reviewer responses. After using this tool, the authors reviewed and edited the content as needed and take full responsibility for the content of the published article.